\documentclass[twocolumn,superscriptaddress,floatfix,preprintnumbers,prl,nofootinbib]{revtex4-2}

\usepackage[colorlinks=true,breaklinks=true]{hyperref}
\usepackage[utf8]{inputenc}

\hypersetup{urlcolor=Blue,
	    citecolor=Blue,
	    linkcolor=Blue}
\usepackage{mathtools}
\usepackage[dvipsnames]{xcolor}

\usepackage{orcidlink}

\newcommand{\unitspace}{\ensuremath{\mskip\thinmuskip}}

\newcommand{\gccm}{\ensuremath{{\mathrm{g}\unitspace\mathrm{cm}^{-3}}}}

\newcommand{\nue}{\ensuremath{\nu_{e}}}
\newcommand{\nuebar}{\ensuremath{\bar{\nu}_{e}}}
\newcommand{\nux}{\ensuremath{\nu_{x}}}
\newcommand{\EtoX}{\ensuremath{\nue,\nuebar \to \nux} }
\newcommand{\XtoE}{\ensuremath{\nux \to\nue,\nuebar} }

\long\def\exclude#1{}

\bibpunct{[}{]}{,}{n}{}{,}

\begin{document}

\title{Fast Neutrino Flavor Conversions can Help and Hinder Neutrino-Driven Explosions}

\author{Jakob Ehring \orcidlink{0000-0003-2912-9978}}
\affiliation{Max-Planck-Institut f\"ur Physik (Werner-Heisenberg-Institut), F\"ohringer Ring 6, D-80805 M\"unchen, Germany}
\affiliation{Max-Planck-Institut f\"ur Astrophysik, Karl-Schwarzschild-Str.~1,
D-85748 Garching, Germany}
\affiliation{Technical University of Munich, TUM School of Natural Sciences, Physics Department, James-Franck-Stra{\ss}e 1, D-85748 Garching, Germany}

\author{Sajad Abbar \orcidlink{0000-0001-8276-997X}}
\affiliation{Max-Planck-Institut f\"ur Physik (Werner-Heisenberg-Institut), F\"ohringer Ring 6,
D-80805 M\"unchen, Germany}

\author{Hans-Thomas Janka \orcidlink{0000-0002-0831-3330}}
\affiliation{Max-Planck-Institut f\"ur Astrophysik, Karl-Schwarzschild-Str.~1,
D-85748 Garching, Germany}

\author{Georg Raffelt \orcidlink{0000-0002-0199-9560}}
\affiliation{Max-Planck-Institut f\"ur Physik (Werner-Heisenberg-Institut), F\"ohringer Ring 6,
D-80805 M\"unchen, Germany}

\author{Irene Tamborra \orcidlink{0000-0001-7449-104X}}
\affiliation{Niels Bohr International Academy \& DARK, Niels Bohr Institute, University of Copenhagen, Blegdamsvej 17, DK-2100 Copenhagen, Denmark}

\date{\today}

%==========================

\begin{abstract}
We present the first simulations of core-collapse supernovae in axial symmetry with feedback from fast neutrino flavor conversion (FFC). Our schematic treatment of FFCs assumes instantaneous flavor equilibration under the constraint of lepton-number conservation individually for each flavor. Systematically varying the spatial domain where FFCs are assumed to occur, we find that they facilitate SN explosions in low-mass (9--12\,M$_\odot$) progenitors that otherwise explode with longer time delays, whereas FFCs weaken the tendency to explode of higher-mass (around 20\,M$_\odot$) progenitors.
\end{abstract}
\maketitle

{\bf\em Introduction and Motivation.}---Multi-dimensional simulations of core-collapse supernovae (CCSNe) with refined, energy-dependent neutrino transport, in particular also in three dimensions (3D), support the viability of the neutrino-driven explosion mechanism \cite{Janka:2012wk, Janka:2016fox, Mueller2016, Janka2017Handbooka, Mueller2020, Burrows+2020, Mezzacappa+2020, Burrows+2021}. Since the huge binding energy of several $10^{53}$\,erg of a newly formed proto-neutron star (PNS) is carried away by neutrinos and antineutrinos of all flavors, about one percent of this energy is well sufficient to explain the vast majority of the observed CCSNe~\citep{Colgate+1966}. The energy transfer to the explosion is mostly mediated by the absorption of several percent of the escaping electron (anti)neutrinos ($\nue$~and $\nuebar$) on free neutrons and protons behind the stalled core-bounce shock~\citep{Bethe+1985}. These reactions therefore decide about success or failure of the explosion. They also determine the neutron-to-proton ratio and thus the formation of chemical elements in the innermost ejecta of successful explosions. Neutrinos of other flavors (commonly pooled as~$\nux$) do not interact via charged-current processes in the ejected matter and thus are subdominant for powering the explosion, although their creation by thermal pair processes contributes significantly to PNS cooling. For numerical models to be predictive, and to understand the processes that trigger and power the blast wave, a reliable implementation of neutrino transport and flavor evolution in the deep SN interior is indispensable~\cite{Duan:2010bg, Mirizzi:2015eza, Tamborra:2020cul, Richers2022_1, Volpe2023_1}.

Neutrino flavor conversions in the SN core remain one of the major uncertainties for rigorous, self-consistent ab-initio modeling. Despite the long-standing insight that neutrinos can change flavor during propagation, the possible consequences have usually been ignored in SN simulations. But the extremely high neutrino number densities in this environment facilitate nonlinear, collective phenomena that are not suppressed by matter effects \cite{Pantaleone1992b}. In particular, $\nu$ and $\bar\nu$ can pairwise undergo so-called fast flavor conversions (FFCs) \cite{Sawyer:2005jk, Sawyer2009a, Sawyer2016a, Izaguirre:2016gsx, Chakraborty:2016lct, Morinaga2022a}. The characteristic length scales depend on the neutrino number density and can be as short as centimeters. FFCs have been at the focus of numerous recent investigations, and their possible occurrence has been diagnosed in different regions inside the PNS, in the neutrino-decoupling layer, and ahead and behind the SN shock \cite{Morinaga:2019wsv, Abbar:2020qpi, Glas2020a, Nagakura:2021hyb, Capozzi:2020syn, Abbar:2018shq}.

Here we explore, for the first time, the impact of FFCs on the evolution of collapsing stars by multi-dimensional neutrino-hyrodynamic simulations, thus expanding our previous work in spherical symmetry \cite{Ehring2023a}. Considering several, distinctly different progenitor models, we demonstrate that the consequences of FFCs depend on the stellar core structure and mass range. The quantum kinetic problem of neutrino flavor transport still awaits a practical solution. Therefore, once more we apply our schematic treatment~\cite{Ehring2023a}, i.e., we assume that below a chosen critical density $\rho_{\rm c}$, FFCs lead to flavor equilibrium under the constraint of lepton-number conservation (thus maximizing the impact of FFCs).

Interestingly, our results show that FFCs can both support or suppress neutrino-driven explosions, with the exact dynamical response depending on the progenitor and the assumed region of flavor conversions.

%---------------------------------------------------------------
\begin{figure*}
        \includegraphics{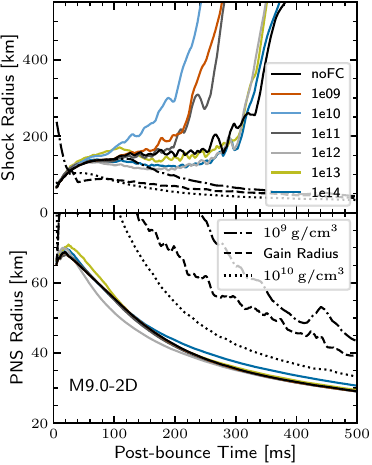}\kern-1pt
        \includegraphics{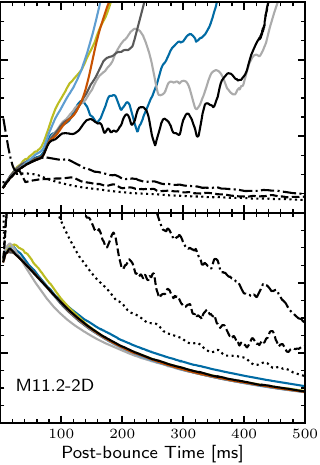}\kern-1pt
        \includegraphics{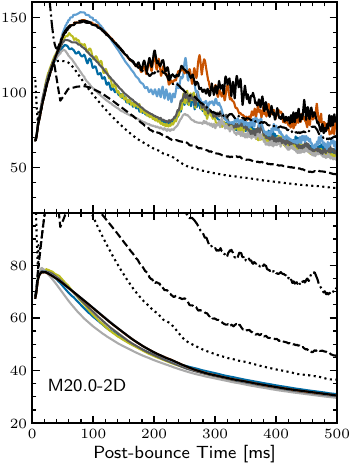}\kern-1.5pt
    \caption{Angle-averaged shock radii (top) and PNS radii (bottom; defined at $\rho=10^{11}\,\gccm$) vs.\ post-bounce time for the indicated models. {\em Black solid lines:} Models noFC (no flavor conversions). {\em Colored solid lines:} Instantaneous FFCs for $\rho<\rho_\mathrm{c}$ as labelled in the legend. The unsteady motion of the average shocks with contraction and expansion phases is caused by the violent large-scale convective mass flows in the neutrino heated gain layer behind the CCSN shock. The sudden growth of $r_{\rm shock}$ (small at $\sim$100\,ms for M9.0-2D and prominent at $\sim$70\,ms for M11.2-2D and at $\sim$220\,ms for M20.0-2D) signals a decrease of mass accretion rate due to the arrival of the Si/O interface. For the noFC models, we also show the angle-averaged gain radius (dashed black) and the mean radii for $\rho=10^9$ and $10^{10}\,\gccm$ (dash-dotted and dotted black lines lines, respectively), all smoothed with 10\,ms running averages. For the 9.0 and $11.2\,M_\odot$ progenitors, FFCs support an earlier onset of the explosion, whereas for $20.0\,M_\odot$ they thwart it and the shock recedes even more rapidly.
    }
    \label{figPRL:Radii}
    \vskip-6pt
\end{figure*}
%---------------------------------------------------------------

{\bf\em Numerical Setup and Set of Simulations.}---Our simulations were performed in axial symmetry (2D) with the neutrino-hydrodynamics code \textsc{Alcar} \cite{2015MNRAS.453.3386J, 2018MNRAS.481.4786J}. This is a state-of-the-art, Eulerian, conservative, higher-order Godunov-type finite-volume solver for the 1D and multi-D fluid dynamics equations coupled to a two-moment scheme for energy-dependent three-flavor neutrino transport. It uses a well-tested implementation of all relevant neutrino processes. The main features, input physics, numerical setup of the models, and description of the FFC implementation are provided in our earlier paper~\cite{Ehring2023a}.

In the absence of a fundamental prescription, and to enable a systematic parametric study, we assume that FFCs take place in the entire volume where $\rho<\rho_\mathrm{c}$, a chosen density threshold value. We assume that pairwise $\nu\bar\nu$ flavor conversion is ``instantaneous,'' i.e., on length scales much less than our numerical cells and time scales defined by our computational time stepping. This approach is justified for FFCs (and any other flavor conversion phenomenon) that proceed on scales much below the resolution of full-scale hydrodynamic CCSN models. We further assume that FFCs lead to complete flavor equilibrium under the constraints of lepton number conservation for each flavor individually, in particular also of electron-neutrino lepton number, as well as energy and total momentum conservation, and with respecting the Pauli exclusion principle. Our algorithm, defined in Eqs.~(9), (10), (14), and (15) of Ref.~\cite{Ehring2023a}, is applied after each time step in each spatial cell where $\rho < \rho_\mathrm{c}$. Some recent studies have focused on the asymptotic FFC state \cite{Zaizen2023a}. We stress that our recipe leads to a converged state: it does not change if the algorithm is applied twice.

Our simulations were evolved in 1D until 5\,ms pb (post bounce) and then mapped onto a 2D polar coordinate grid consisting of 640 logarithmically spaced radial zones and 80 equidistant lateral ones. The central 2\,km core was still calculated in 1D, permitting larger time steps, yet having negligible influence on the hydrodynamic evolution. During the mapping, random cell-by-cell perturbations of 0.1\,\% of the local density were applied to seed the hydrodynamic instabilities, which otherwise would develop only due to uncontrolled numerical noise.

We selected three progenitors with different zero-age main-sequence masses. One is the 20\,M$_\odot$ model \cite{2007PhR...442..269W} that we used in our previous 1D study \cite{Ehring2023a}. In addition, we investigated a 9\,M$_\odot$ \cite{Woosley+2015} and 11.2\,M$_\odot$ model \cite{Woosley+2002}. The 9\,M$_\odot$ star consistently explodes in multi-D simulations, although in some more quickly and about twice as energetically \cite{Radice+2017,Burrows+2020,Burrows+2021} than in others \cite{2018MNRAS.481.4786J,Glas+2019,Stockinger+2020}. The 11.2\,M$_\odot$ model is less ready to blow up, exhibiting a delayed and slow onset of shock expansion \citep{Buras+2006,Marek+2009,Takiwaki+2012,Mueller2015}. In contrast, the 20\,M$_\odot$ star failed to explode in most multi-D simulations \cite{Melson+2015,2018MNRAS.481.4786J,Vartanyan+2018,Glas+2019}.

The convention for naming our simulations follows our previous one \cite{Ehring2023a}, supplemented with a numerical value for the stellar mass: M9.0-2D-xxx, M11.2-2D-xxx, and M20.0-2D-xxx. Here xxx is a placeholder for either noFC (``no flavor conversion'') or for the FFC threshold density. We implement $\rho_\mathrm{c}=10^9$\,g\,cm$^{-3}$, ..., $10^{14}$\,g\,cm$^{-3}$ in steps of factors of 10, corresponding to xxx = 1e09, ..., 1e14.

%-----------------------------------------------------------------
\begin{figure}
    \includegraphics{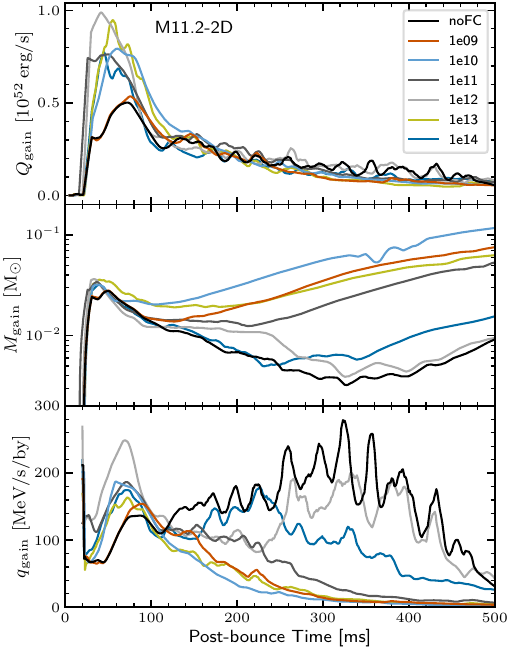}
    \caption{
    Total net neutrino-heating rate (top), mass in the gain layer (middle), and net heating rate per baryon (bottom) for our M11.2-2D simulations. (10\,ms running average.)
    }
    \label{figPRL:Heating_S11.2_2D}
\end{figure}
%-------------------------------------------------------------------

%---------------------------------------------------------------
\begin{figure*}
    \includegraphics{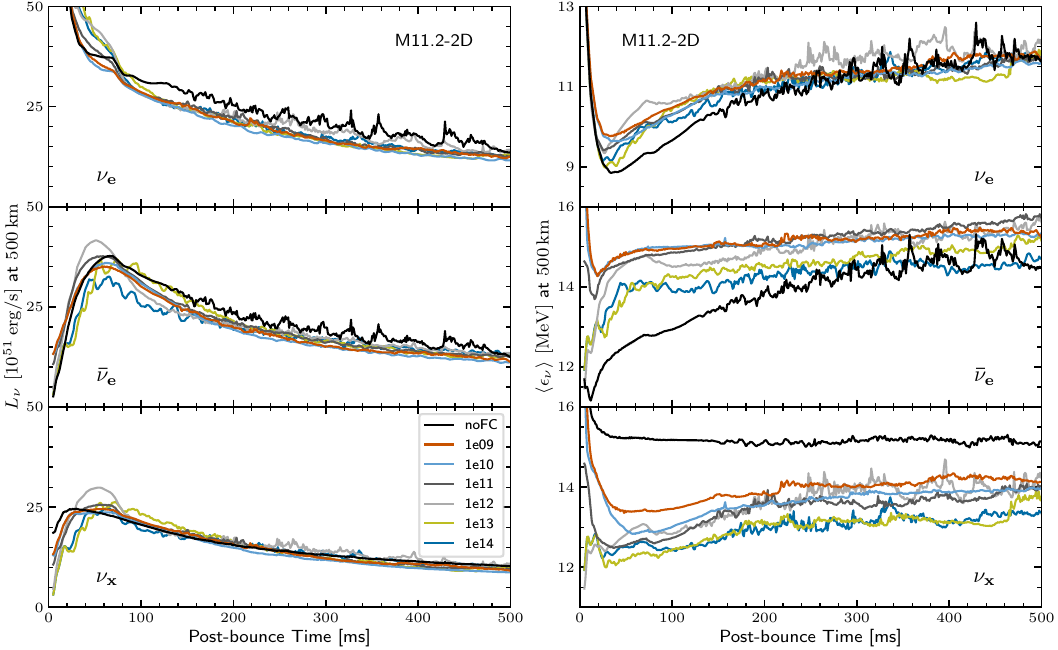}
    \vskip-5pt
    \caption{
    Evolution of luminosities and mean energies of the radiated $\nu_e$, $\bar\nu_e$, and one species of $\nu_x$ for our M11.2-2D simulations.
    All quantities are angle-averaged, measured at a radius of 500\,km, and transformed to a distant observer at rest.
    }
    \label{figPRL:Neutrinos_S11.2_2D}
    \vskip-5pt
\end{figure*}
%---------------------------------------------------------------

%-----------------------------------------------------------
\begin{figure}
    \includegraphics{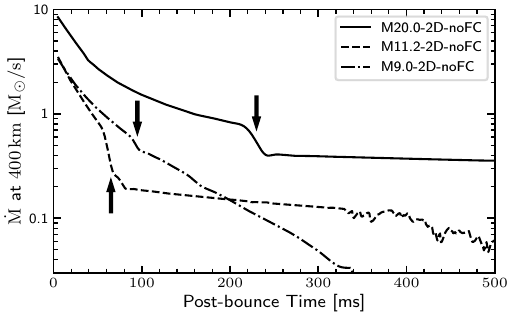}
    \caption{Mass accretion rates at 400\,km for the noFC models.
    The step-like features (arrows) derive from density jumps at the Si/O interfaces of the collapsing progenitors.
    }
    \label{figPRL:MassAccretionRates}
\end{figure}
%-----------------------------------------------------------

{\bf\em Results.}---In our previous 1D simulations \cite{Ehring2023a} of the 20\,M$_\odot$ progenitor we found that FFCs caused a faster and stronger shock contraction than without FFCs for all threshold densities $\rho_\mathrm{c}$ and for all times (except for $\rho_\mathrm{c} = 10^{10}$\,g\,cm$^{-3}$ during a short period of about 70\,ms around 100\,ms pb). This finding suggested that FFCs tend to hinder shock revival and neutrino-driven explosions, and this conclusion is confirmed in 2D for the 20\,M$_\odot$ star (Fig.~\ref{figPRL:Radii}).

However, our 9 and 11.2\,M$_\odot$ progenitors demonstrate that this is not generally the case (Fig.~\ref{figPRL:Radii}). Including FFCs, in particular for $\rho_\mathrm{c} = 10^9$, $10^{10}$, $10^{11}$\,g\,cm$^{-3}$, yields significantly earlier explosions. The main explanation is a higher net heating rate per nucleon ($q_\mathrm{gain}$) for at least $\sim$100\,ms pb. The increased $q_\mathrm{gain}$ causes a persistently higher mass $M_\mathrm{gain}$ in the energy-gain layer behind the stalled shock and a correspondingly higher total net heating rate $Q_\mathrm{gain}$ in this volume (Fig.~\ref{figPRL:Heating_S11.2_2D}).

This enhanced heating can be traced to \XtoE conversions of high-energy heavy-lepton neutrinos exterior to the neutrinospheres. Although the $\nu_e$ and $\bar\nu_e$ number fluxes exceed the $\nu_x$ ones during the accretion phase before the onset of the explosion, equilibration with the more energetic $\nu_x$ raises the mean $\nu_e$ and $\bar\nu_e$ energies, $\langle\epsilon_{\nue}\rangle$ and $\langle\epsilon_{\nuebar}\rangle$ (Fig.~\ref{figPRL:Neutrinos_S11.2_2D}, right). Despite the concomitant decrease of the $\nu_e$ and $\bar\nu_e$ luminosities (Fig.~\ref{figPRL:Neutrinos_S11.2_2D}, left), the energy transfer to the postshock layer is boosted because $q_{\rm gain}\propto\langle\epsilon_\nu^2\rangle$.

The set of simulations M9.0-2D-1e09, 10, 11 as well as M11.2-2D-1e09, 10, 11 each exhibit a similarly fast expansion of their shock trajectories. In particular, even the $\rho_\mathrm{c} = 10^9$\,g\,cm$^{-3}$ cases feature such a rapid shock revival because the surface of constant $\rho=10^9$\,g\,cm$^{-3}$ moves behind the shock (Fig.~\ref{figPRL:Radii}, dash-dotted black lines). Therefore \XtoE conversions take place in a major part of the gain layer even for this low conversion density.

The FFC impact on the post-bounce evolution changes when flavor equilibration occurs deeper inside. In M9.0-2D-1e12 and M11.2-2D-1e12, the explosion sets in at nearly the same time as without FFC, although the shock transiently expands to nearly 500\,km at $\sim$220\,ms pb in M11.2-2D-1e12. The subsequent breakdown of the shock expansion follows from extremely rapid PNS contraction (Fig.~\ref{figPRL:Radii}), caused by enhanced cooling due to \EtoX conversions in the layer between $10^{11}$ and $10^{12}$\,g\,cm$^{-3}$. The $\nu_x$ can easily escape from this region exterior to their neutrinosphere. The corresponding luminosities increase visibly in Fig.~\ref{figPRL:Neutrinos_S11.2_2D} not only for $\nu_x$, but also for $\bar\nu_e$ (and slightly for $\nu_e$) because of the compression heating of their neutrinospheres. This fast PNS contraction also pulls matter from the gain layer into the cooling layer, whose mass is lower than in the quickly exploding model M11.2-2D-1e11 at all times. After $\sim$250\,ms,  $q_{\rm gain}$, $M_{\rm gain}$ and  $Q_{\rm gain}$ of M11.2-2D-1e12 become similar to noFC (Fig.~\ref{figPRL:Heating_S11.2_2D}), for which reason both models ultimately explode in a synchronous way. Similar arguments linked to the faster PNS contraction also apply for M9.0-2D-1e12.

In M11.2-2D-1e14,  $q_{\rm gain}$ and  $Q_{\rm gain}$ before $\sim$100\,ms are higher than in M11.2-2D-noFC, but stronger cooling between the gain radius $r_\mathrm{gain}$ and PNS radius $r_\mathrm{PNS}$ (at $10^{11}$\,g\,cm$^{-3}$) prevents a fast explosion of this model. After $\sim$100\,ms heating and cooling in the gain and cooling layers and $M_{\rm gain}$ differ little between the two models (Fig.~\ref{figPRL:Heating_S11.2_2D}). Therefore M11.2-2D-1e14 explodes only slightly earlier, possibly triggered by stochastically intensified postshock convection. M9.0-2D-1e13 and 14 compared to noFC show amplified cooling below $r_{\rm gain}$ for roughly 150\,ms, and smaller $M_{\rm gain}$ and $Q_{\rm gain}$ from this time until twice longer. In all of these models, including M9.0-2D-1e12, the explosion sets in only later than $\sim$300\,ms (Fig.~\ref{figPRL:Radii}).

M11.2-2D-1e13 is a stark outlier: it sports rapid shock expansion very early and similar to M11.2-2D-1e10, but in clear contrast to M9.0-2D-1e13. The favorable explosion conditions in M11.2-2D-1e13 derive from $r_{\rm PNS}$ exceeding that of any other simulation of this star during the first 100\,ms (Fig.~\ref{figPRL:Radii}). This permits more mass to stay in the gain layer during this period when $q_{\rm gain}$ is enhanced by FFC effects (Fig.~\ref{figPRL:Heating_S11.2_2D}). These conditions facilitate the runaway shock expansion when the composition interface between the progenitor's Si shell and Si-enriched O-shell reaches the shock at $t_{\rm pb}\sim70$\,ms. The larger $r_{\rm PNS}$ is caused by very strong PNS convection, which is boosted by local heating around $\rho_\mathrm{c} = 10^{13}$\,g\,cm$^{-3}$. This happens when $\nu_x$ diffusing out from the PNS core are converted to $\nu_e$ and $\bar\nu_e$ just below this threshold density and the newly created $\nu_e$ and $\bar\nu_e$ get rapidly absorbed in the local medium (see \cite{Ehring2023a} for a detailed description of this effect in 1D). This intriguing convective boost is much weaker when \XtoE happens at $\rho_\mathrm{c} = 10^{14}$\,g\,cm$^{-3}$.

This FFC effect on PNS convection and increased $r_{\rm PNS}$ also appears in M9.0-2D-1e13 (Fig.~\ref{figPRL:Radii}). However, in this model the mass accretion rate $\dot M$ decreases gradually and a pronounced decline of the density and thus of the mass accretion rate at the progenitor's Si/O interface are absent (Fig.~\ref{figPRL:MassAccretionRates}). $\dot M$ in the 9.0\,M$_\odot$ star drops below the one in 11.2\,M$_\odot$ only at $t_{\rm pb}\sim 200$\,ms. At this time, $q_{\rm gain}$, $M_{\rm gain}$ and  $Q_{\rm gain}$ of M9.0-2D-1e13 have come close to noFC and therefore both models exhibit a very similar later shock evolution.

Our findings are completely different for the 20\,M$_\odot$ progenitor, where FFCs do not improve the explosion conditions. Without FFCs, there is no explosion within 500\,ms pb, and in all FFC cases, the average $r_{\rm shock}$ is even smaller most of the time, with only a few short inversions (Fig.~\ref{figPRL:Radii}). The overall shock evolution resembles our 1D simulations \cite{Ehring2023a}, except for some secondary differences of the PNS contraction and a corresponding reaction of the shock due to the FFC-induced boosting of the PNS convection in M20.0-2D-1e13 and 14. All models fail to explode, because at the time ($\sim$230\,ms) when the Si/O composition interface arrives at the shock, $q_{\rm gain}$, $M_{\rm gain}$ and $r_{\rm shock}$ in M20.0-2D-1e09 are close to noFC and in all other FFC models considerably lower.

In this critical phase, the weaker FFC-implied $q_{\rm gain}$ is traced to the high mass infall rate in the collapsing 20\,M$_\odot$ star. This leads to the formation of a very massive PNS ($\sim$1.9\,M$_\odot$ instead of 1.30--1.36\,M$_\odot$ in the lower-mass progenitors) that accumulates a hot accretion mantle. As this PNS contracts, $\langle\epsilon_{\nue}\rangle$ and $\langle\epsilon_{\nuebar}\rangle$ increase quickly in M20.0-2D-noFC, whereas $\langle\epsilon_{\nu_x}\rangle$ remains nearly constant. Therefore \XtoE can result in only moderately higher $\langle\epsilon_{\nue}\rangle$ and $\langle\epsilon_{\nuebar}\rangle$. In $q_{\rm gain}$, these marginally higher $\langle\epsilon_{\nue}\rangle$ and $\langle\epsilon_{\nuebar}\rangle$ cannot compensate for the considerable reduction of the $\nu_e$ and $\bar\nu_e$ luminosities caused by flavor equilibration.

Whether FFCs help or hinder the CCSN explosion is therefore  tightly connected to $\dot M$ and the corresponding growth of the PNS mass and $\langle\epsilon_\nu\rangle$ of the radiated neutrinos. This suggests that the dynamical impact of FFCs correlates with the compactness defined in Ref.~\cite{OConnor+2011} for characterizing the stellar core structure.

{\bf\em Discussion.}---We have studied the possible impact of FFCs on CCSNe by 2D simulations, using a schematic treatment that enforces instantaneous (i.e., on subgrid scales) flavor equilibration under the constraints of energy, momentum, and individual lepton number conservation. We have systematically varied the threshold density below which FFCs are assumed to occur. 

Our results suggest that FFCs can both facilitate or weaken the onset of runaway shock expansion, depending on the core structure of the stellar progenitor and the region where FFCs are assumed to occur. Concerning their supportive influence on explosions, FFCs are on a par with a variety of other effects that assist the neutrino-driven mechanism. In particular the 3D nature of pre-collapse stars in terms of density and velocity perturbations associated with convective oxygen and silicon shell burning is important for the successful shock revival in 3D simulations \citep{Mueller+2017,Bollig+2021}. Neutrino-powered explosions are also fostered by other effects such as muons in the high-density medium \citep{Bollig:2017lki,Bollig+2021}, strangeness-dependent contributions to the axial-vector coupling constant of neutral current neutrino-nucleon scattering~\citep{Melson+2015,Horowitz+2017}, a higher effective nucleon mass at densities above roughly 10\,\% of the nuclear saturation density \cite{Schneider+2019,Yasin+2020}, or magnetic fields. The B-fields can aid the initiation of the shock expansion even without the field-amplifying effects of rapid rotation, provided the pre-collapse core of the progenitor star is strongly magnetized \citep{Obergaulinger+2014,Mueller+2020,Matsumoto+2022,Varma+2023}.

Also non-standard physics such as a hadron-quark phase transition~\citep{Fischer+2018}, beyond-standard-model particle physics \citep{Rembiasz+2018,Mori+2022}, and modified theories of gravity~\citep{Kuroda+2023} have been suggested as potentially supportive to shock revival. In contrast to such possibilities, which reach beyond the limits of currently well constrained physics, FFC is a phenomenon that occurs within the framework of well established theory, and its consequences for CCSNe should therefore be better understood.

Further work on practical solutions of the neutrino quantum kinetic equations in SNe and neutron star (NS) mergers is therefore imperative~\cite{Nagakura2023b}. And simulations for a wider range of progenitors and longer evolution times are needed to explore the consequences of FFCs on observable signals and CCSN properties, for example explosion energies, nucleosynthesis in neutrino-heated ejecta, neutrino signals from all evolution phases, and gravitational waves, which could be amplified by stronger PNS convection.

In contrast to most of the other effects mentioned above, we found the impact of FFCs on the CCSN dynamics to differ in low-mass and high-mass progenitors. This is an intriguing result and therefore requires verification by more elaborate treatments of the neutrino flavor evolution to replace our simplifying assumptions.

If our results are confirmed, they could also be relevant for the mass distributions of NSs and black holes (BHs). Earlier explosions due to FFCs can lower the minimum NS mass expected from stellar collapse and might help to explain the formation of NSs below 1.2\,M$_\odot$, whose observation challenges current CCSN models  \cite{Stockinger+2020,Suwa+2018}. For example, our model M11.2-2D-1e10 produces a NS with a baryonic mass of only 1.30\,M$_\odot$ (gravitational mass of $\sim$1.19\,M$_\odot$ for 12\,km radius; \cite{Lattimer+2001}) instead of $\sim$1.36\,M$_\odot$ without FFC. On the other hand, more difficult explosions for massive progenitors would not only raise the BH formation rate, but could also bear on the red supergiant problem, i.e., the lack of observed type-II SNe from progenitors with zero-age-main-sequence masses above 17--20\,M$_\odot$  \cite{Smartt+2009}, although current 2D and 3D models yield explosions well beyond this mass range \cite{Nakamura+2015,Burrows+2021}.

{\bf\em Acknowledgments.}---J.E.\ is grateful to Robert Glas and Oliver Just for their introduction to the use of the \textsc{Alcar} code.
This work was supported by the German Research Foundation (DFG) through the Collaborative Research Centre ``Neutrinos and Dark Matter in Astro- and Particle Physics (NDM),'' Grant No.\ SFB-1258-283604770, and under Germany’s Excellence Strategy through the Cluster of Excellence ORIGINS EXC-2094-390783311. In Copenhagen this project received funding from the Villum Foundation (Project No.~37358) and the Danmarks Frie Forskningsfonds (Project No.\ 8049-00038B).
SA, GR, and HTJ would like to express special thanks to the Mainz Institute for Theoretical Physics (MITP) of the Cluster of Excellence PRISMA+ (Project ID~39083149), for its hospitality and its partial support during the completion of this work.
We also acknowledge the use of the following software: \textsc{Matplotlib}~\cite{Matplotlib}, \textsc{Numpy}~\cite{Numpy}, \textsc{SciPy}~\cite{SciPy}, \textsc{IPython}~\cite{IPython}.

\bigskip

\bibliographystyle{bibi}
\bibliography{references}

\end{document}